\documentclass[11]{article}

\usepackage[dvips]{graphicx}
\newcommand{\vx}{\mbox{\boldmath $x$}}

\newcommand{\vz}{\mbox{\boldmath $z$}}

\newcommand{\vp}{\mbox{\boldmath $p$}}

\begin{document}

\title{\bf Time-Dependent Dynamics of the Bose-Fermi \\
Mixed Condensed System
\footnote{Proc. of Int. Conf. CM2002}}

\author{T. Maruyama$^{a,b}$, H. Yabu$^{b}$ 
and T. Suzuki$^{b}$ \\
a) College of Bioresource Sciences, Nihon University, Japan\\
b) Department of Physics, Tokyo Metropolitan University, Japan}


\maketitle

\begin{abstract}
We study the monopole oscillation in the bose-fermi mixed condensed
system by performing the time-dependent Gross-Pitaevsky (GP) and  Vlasov
equations.
We find that the big damping exists for the fermion oscillation
in the mized system even at zero temperature  
\end{abstract}

It is one of the most excited theme in recent physics to study
the time-dependent dynamical motions of traped atoms
under the existence of  the Bose-Einstein condensates (BEC) \cite{bec,Dalfovo}, 
such as the collective motions\cite{osc}, quantum votices \cite{vortex} and 
atomic novae \cite{colap}.
In the view of the theoretical research, these phenomena are
very important also to construct and to examine 
the transport theory in finite many body system. 
In fact  many atomic system must give good probes
for such study because
the fundamental interaction is clear and weak.

In this work we construct the transport model including the condensed bosons 
and fermions, and as a first step study the collective monopole motion.

Here we briefly explain our formalism.
First we define the Hamiltonian for boson-fermion coexistent system
as follows.
\begin{equation}
H = H_B + H_F + H_{BF}
\end{equation}
with
\begin{eqnarray}
H_B &=& \int d^3 x \{ - \frac{1}{2} \phi^{\dagger}(\vx) \nabla^2 \phi(\vx)
+ \frac{1}{2} \vx^2 \phi^{\dagger}(\vx) \phi(\vx)
+ \frac{g_B}{2} (\phi^{\dagger} \phi)^2 \} ,
\\
H_F &=& \int d^3 x \{ - \frac{\hbar^2}{2m_f} \psi^{\dagger} \nabla^2 \psi
+ \frac{1}{2} m_f \omega_f^2 \psi^{\dagger} \psi \} ,
\\
H_{BF} &=& h_{BF} \int d^3 x \{ \phi^{\dagger} \phi \psi^{\dagger} \psi \} ,
\end{eqnarray}
where $\phi$ and $\psi$ are boson and fermion fields, respectively.
Fermion mass $m_f$ and the trapped frequency $\omega_f$ are normalized
with the boson mass $M_B$ and the boson trapped frequency $\Omega_B$, respectively.
The coordinates are normalized by $\xi_B = (\hbar / M_B \Omega _B)^{1/2}$.
The coupling constants $g_B$ and $h_{BF}$ are given as
\begin{eqnarray}
g_B &=&  4 \pi a_{BB} \xi_B^{-1},  \\
h_{BF} &=&  2 \pi a_{BF} \xi_B^{-1} (1 + m_f^{-1}),
\end{eqnarray} 
where $a_{BB}$ and  $a_{BF}$
are the scattering lengths between two bosons
and between the boson and the fermion, respectively.

The total wave function ${|\Phi(\tau)>}$ has  $N_c$ condensed bosons, whose
wave function $\phi_c$ is defined as
\begin{equation}
\phi_c (\vx,\tau) = <\Phi|\phi_B(\vx,\tau)|\Phi>,
\end{equation}
where $\tau$ is the time coordinate normalized with $\Omega_B^{-1}$.
In this work the wave function $\phi_c$ is expanded 
with the harmonic oscillator wave function $u_n(\vx)$ as
\begin{eqnarray}
&&\phi_c (\vx,\tau) = \sum_{n=0}^{N_{base}} A_n e^{i \theta_n} u_n(b\vx^2) 
e^{-\frac{i}{2} \nu\vx^2} , 
\label{c-tran}
\end{eqnarray}
where $N_{base}+1$ is the number of the harmonic oscillator bases.
We define the Lagrangian with the collective coordinates as
\begin{equation}
L(A_n, \theta_n, b, \nu) = 
< \Phi(\tau)| \{ i \frac{\partial}{\partial \tau} - H \} | \Phi (\tau) >.
\end{equation}
Instead of solving the time-dependent GP equation directly,
we take $A_n, \theta_n, b, \nu$ as time-dependent variables 
and solve the Euler-Lagrange equations with respect to
these variables. 

The many fermion system can be described with the Thomas-Fermi
approximation, which is given in the limit $\hbar \rightarrow 0$.
Here we define the phase-space distribution function  as
\begin{equation}
f(\vx,\vp,\tau) = \int {d^3 z} <\Phi|
\psi(\vx+\frac{1}{2}{\vz},\tau)
\psi^{\dagger}(\vx-\frac{1}{2}{\vz},\tau) |\Phi>
 e^{-i \vp \vz} .
\end{equation}
In this classical limit
this phase-space distribution function satisfies
the following Vlasov equation:
\begin{equation}
\frac{d}{d \tau} f(\vx,\vp;\tau) =
\{ \frac{\partial}{\partial \tau} + \frac{\vp}{m_f}{\nabla_x} -
 [\nabla_x U_F(x)][\nabla_p] \} f(\vx,\vp;\tau) = 0 .
\label{Vlasov}
\end{equation}
In the actual calculation we solve the above Vlasov equation (\ref{Vlasov}) 
with the test particle method \cite{TP}.

Thus we can get the time-evolution of 
the condensed bosons and the fermions
by solving the time-dependent Gross-Pitaevskii equation
and the Vlasov equation.
As a first step we apply this method to the monopole vibration.

\bigskip

We calculate the monopole vibration
of the system $^{39}$K-$^{40}$K;
the number of the bosons ($^{39}$K) and the fermions ($^{40}$K)
are taken to be 100,000 and 1,000, respectively.
The trapped frequencies are taken as $\Omega_B = 100$ (Hz) and 
$\omega_f = 1$.
Furthermore we use the interaction parameters as
$a_{BB} = 4.22$(nm) and $a_{BF} = 2.51$(nm) \cite{miyakawa}.

Using the root-mean-square radius (RMSR) $R$,  we here define 
\begin{equation}
\Delta r (\tau) = R (\tau) /{R}_0 - 1,
\end{equation}
where  $R_0$~ is the RMSR of the ground state, and $\tau$ is the time
valiable normalized by $\Omega_B^{-1}$.
In Fig.~1 we show  the time-dependence of this observable for bosons
($\Delta r_B$) and fermions ($\Delta r_F$).
In this calculation the initial condition is taken to be
$\Delta r_B = 0$ and  $\Delta r_F = -0.1$.

\begin{center}
\begin{minipage}{15cm}
\vspace*{0.2cm}
\includegraphics[scale=0.7,angle=270]{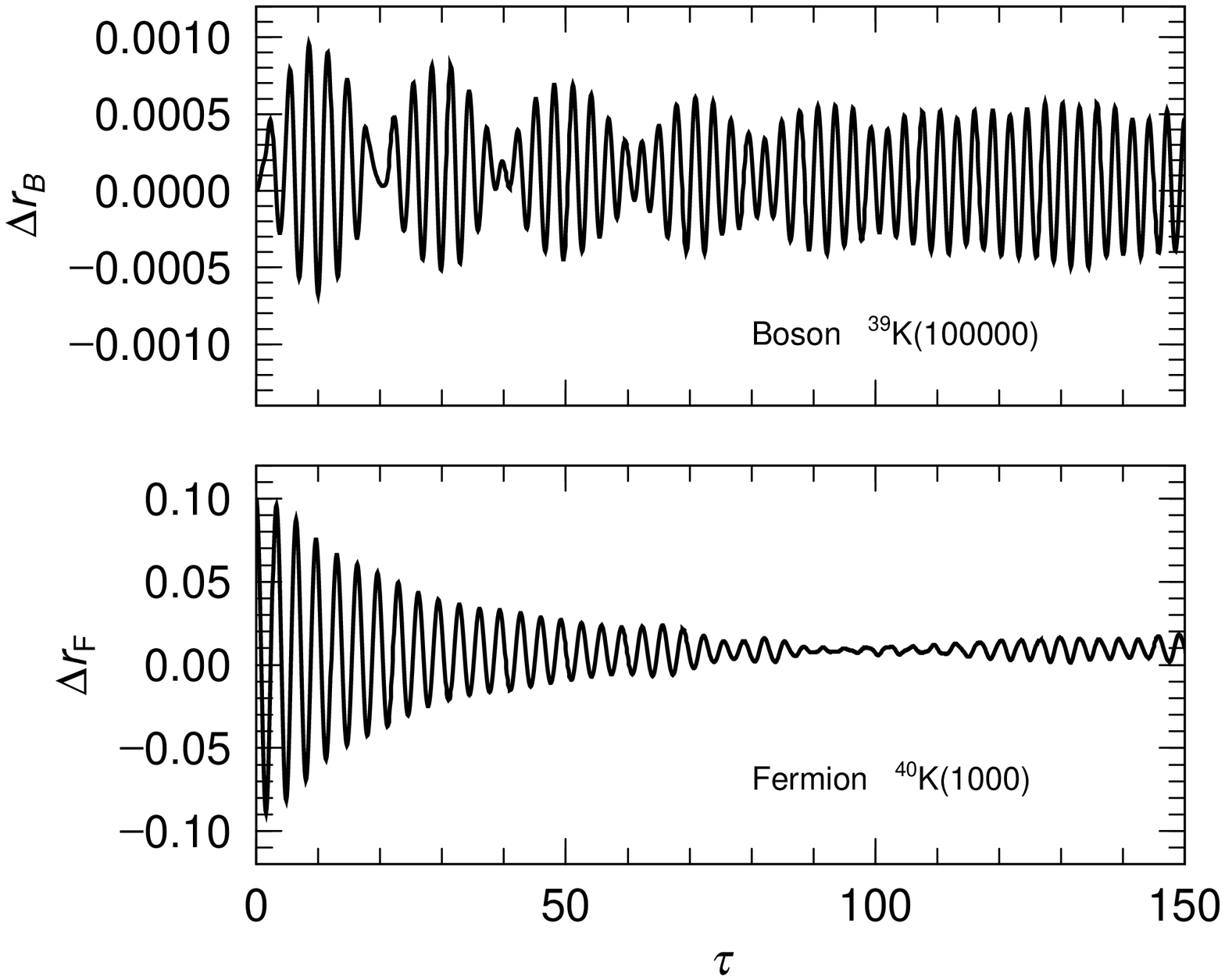}
\medskip

{\small {\bf Fig. 1:}
Time evolution of $\Delta r$ for boson (a) and fermion (b).
}
\end{minipage}
\end{center}

\bigskip

We see a fast damping in the fermion oscillation.
The bosonic oscillation, too, is not harmonic.
The latter should originate from the bose-fermi coupling,
as otherwise the bosonic one is almost daping free at zero temperature.
 \cite{chevy}.

In order to study this phenomenon, futher, we examine the spectrum 
which is obtained by the Fourier transformation
\begin{equation}
F(\omega) = {A_C} |\int d \tau R^2(\tau) e^{i \omega \tau}|^2,
\end{equation}
where $A_C$ is a normalization factor. 
In Fig.~2 we give two typical results
for the spectra obtained by integrating over
the regions $0 < \tau < 60$ (Fig. 2a and 2b)  
and over $100 < \tau < 400$ (Fig. 2c and 2d). 

In Fig. 2a we see that  the boson vibration mainly has two large peaks
at the frequency $\omega = $ 1.91 and 2.21 in the early time stage.
In this time region the fermion oscillation has one peak at $\omega = 1.91$ 
with broad width (Fig. 2b).

Then we can consider the phenomena in the early time stage as follows.
At first stage only the fermions moves, and
this fermion oscillation triggers the boson vibration.
The intrinsic frequencies of the boson and fermion oscillations
are $\omega \approx 2.2$ and 1.9, respectively.
Thus the motion of the bosons corresponds to
the forced vibration, and the beat appears in the bosonic vibration.
Furthermore the boson motion scatters fermions, so that
bosons mediate fermion-fermion interaction
and causes a damping in the fermion vibration. 

In later time stage 
the strength of the boson vibration distributes to 
higher frequency modes 
while the strength for the fermion  almost  concentrates to
one mode with $\omega = 1.91$. 
In addition the beat of the boson vibration disappears
in this later time stage (Fig.~1a).

\bigskip

\begin{center}
\begin{minipage}{15cm}
\vspace*{0.5cm}
\includegraphics[scale=0.6,angle=270]{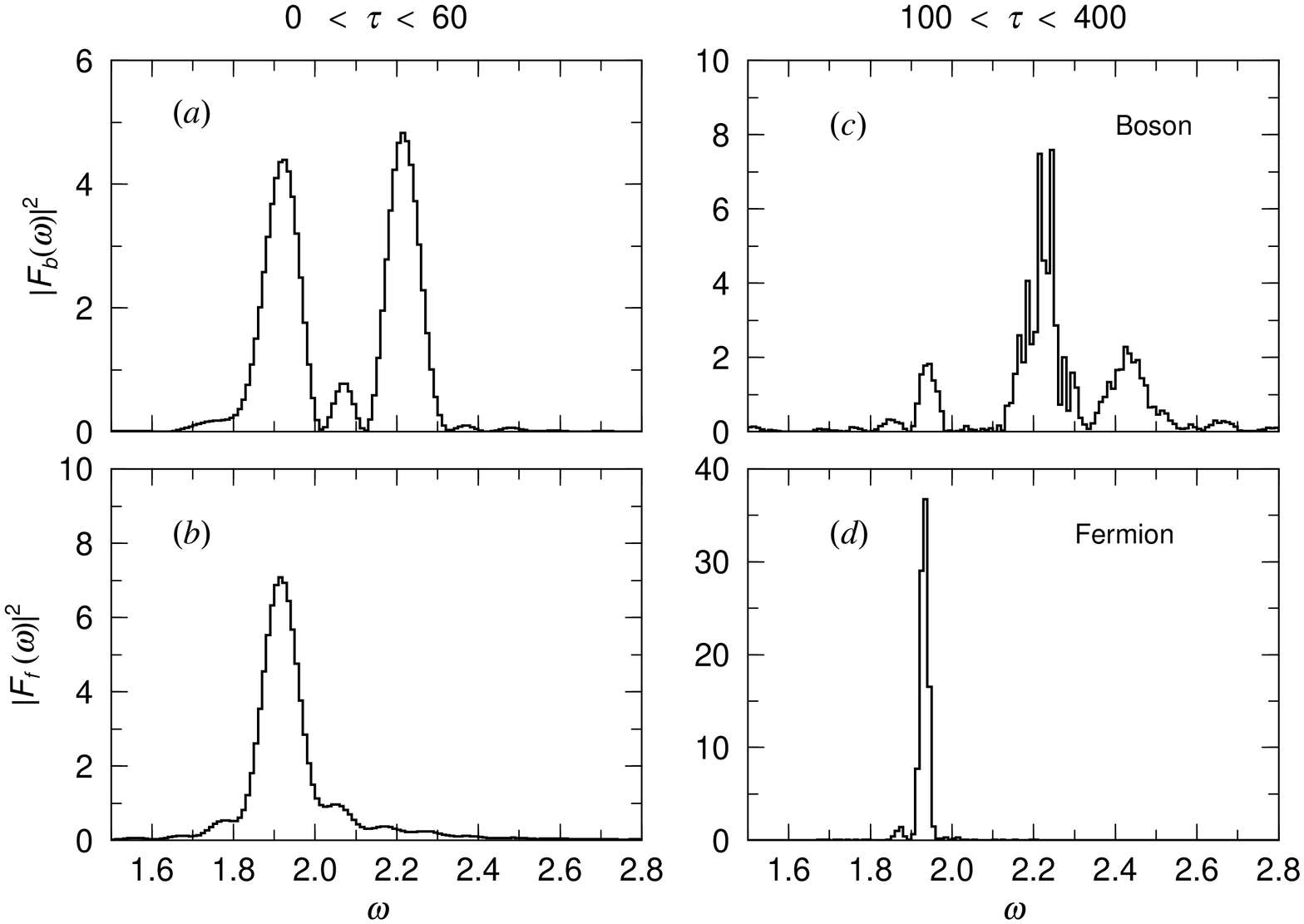}
\medskip

\noindent
{\small {\bf Fig. 2:}
Spectra of the boson oscillation in upper columns (a,c) and
fermion oscillation in down columns (b,d).
The spectra deduced from the evolution in $0 < \tau < 60$ 
are shown in the left columns 
and that in $100 < \tau < 400$ are in the right columns.
}
\end{minipage}
\end{center}

\bigskip

The behavior of this oscillation is never repetitive. 
In the actual calculation the fermion energy $<H_F + H_{BF}>$ does not change  
in the time evolution process. 
The oscillation energy of the fermions isy turned to another kind of energy,
which might be the thermal energy.
We will investigate this possibility more in future.


%
%
%
%


\begin{thebibliography}{0}

\bibitem{bec} M.~H.~Anderson, et al. 
Science {\bf 269}, 198 (1995);\\
K.~B.~Davis, et al., Phys.~Rev.~Lett. {\bf 75}, 3969 (1995).
%
\bibitem{Dalfovo} F.~Dalfovo, et al.,
Rev.~Mod.~Phys. {\bf 71}, 463 (1999).
%
\bibitem{osc} D.~S.~Jin, et al. Phys.~Rev.~Lett. {\bf 77}, 420 (1996);\\
M.-O.~Mewes, et al., Phys.~Rev.~Lett. {\bf 77}, 988 (1996).
%
\bibitem{vortex}
M.R. Matthews at al., Phys. Rev. Lett. {\bf 83}, 2498 (1999).

\bibitem{colap}
J.L. Roberts, et al., Phys. Rev. Lett. {\bf 86}, 4211 (2001). 

\bibitem{miyakawa}
T. Miyakawa et al., J. Phys. Soc. Japan, {Vol. 69} (2000) 2779.

\bibitem{sogo}
T. Sogo, et al., Phys. Rev. {\bf A 66}, 013618 (2002). 

\bibitem{TP}
C.Y. Wong, Phys. Rev. {\bf C25}, 1460 (1982).

\bibitem{chevy}
F. Chevy, Phys. Rev. Lett. {\bf 88}, 250402 (2002). 

\bibitem{Jack1}
B. Jackson and E. Zaembra, cond-mat/0205421.

\end{thebibliography}
\end{document}